\documentclass{elsart}
\usepackage{amsmath}
\usepackage{epsfig}
\usepackage{hyperref}
\usepackage{graphicx}
\usepackage[usenames]{color}
\usepackage[normalem]{ulem}

\begin{document}

\begin{frontmatter}

\title{Engineering Bright Solitons to Enhance the Stability of
Two-Component Bose-Einstein Condensates}
\author{R. Radha$^{\ast,1}$}
\corauth{Corresponding author.}
 \ead{radha\_ramaswamy@yahoo.com, \\
 Telephone: (91)-0435-2403119, Fax: (91)-0435-2403119}
\author{P. S. Vinayagam$^1$}
\author{J. B. Sudharsan$^1$}
\author{Wu-Ming Liu$^{\dagger,2}$}
\ead{wmliu@iphy.ac.cn$^{\dagger}$}
\author{Boris A. Malomed$^{\dagger\dagger,3}$}
\ead{malomed@post.tau.ac.il$^{\dagger \dagger}$}
\address{$^1$ Centre for Nonlinear Science, PG and Research Dept. of Physics,
Govt. College for Women (Autonomous), Kumbakonam 612001, India}
\address{$^2$ Beijing National Laboratory for Condensed Matter Physics,
Institute of Physics, Chinese Academy of Sciences, Beijing-100190,
China.}
\address{$^3$ Department of Physical Electronics, School of Electrical
Engineering, Faculty of Engineering,Tel Aviv University, Tel Aviv
69978, Israel.}

\begin{abstract}
We consider a system of coupled Gross-Pitaevskii (GP) equations
describing a binary quasi-one-dimensional Bose-Einstein condensate
(BEC) with intrinsic time-dependent attractive interactions,
placed in a time-dependent expulsive parabolic potential, in a
special case when the system is integrable (a deformed Manakov's
system). Since  the nonlinearity in the integrable system which
represents binary attractive interactions exponentially decays
with time, solitons are also subject to decay. Nevertheless, it is
shown that the robustness of bright solitons can be enhanced in
this system, making their respective lifetime longer, by matching
the time dependence of the interaction strength (adjusted with the
help of the Feshbach-resonance management) to the time modulation
of the strength of the parabolic potential. The analytical
results, and their stability, are corroborated by numerical
simulations. In particular, we demonstrate that the addition of
random noise does not impact the stability of the solitons.
\end{abstract}
\begin{keyword}
Bose-Einstein condensate; GP equation; Bright Soliton; Gauge transformation\\
{PACS: 05.45.Yv,03.75.-b}
\end{keyword}
\end{frontmatter}\newpage

\section{Introduction}

The experimental realization of Bose-Einstein condensates (BECs)
in dilute gases \cite{bec01} has provided a deep insight into the
realm of macroscopic quantum phenomena, and has now become an
experimental testbed for direct manipulations of matter waves. The
creation of bright \cite{bright02} and dark \cite{dark03} solitons
in BEC has drawn additional strong interest to this area
\cite{ourreview}. The behavior of the single-component (scalar)
BEC is affected by the external trapping potential and the
interatomic collisions, characterized by the scattering length.
Actually, the lifetime of the self-attractive scalar BEC is very
short, as it is subject to decay, once its density exceeds a
critical value. Therefore, effective stabilization of
self-attractive condensates remains an important problem. In this
context, the use of binary (two-component) BECs, composed of
either hyperfine states of the same atom \cite{old,new}, or
different atomic species \cite{twospecies05}, may be very
relevant, as the dynamics of binary condensates can be controlled
by adjusting both the intra- and inter-species scattering lengths
via the Feshbach resonance (FR), see book \cite{book} and
references therein. In particular, the concept of the exchange of
energy between the two components of a binary system, which is
known in the realm of optical solitons \cite{RR:ML:97}, has been
demonstrated in binary condensates as well \cite{Rajendren}. This
mechanism can be employed to increase the lifetime of the
quasi-one-dimensional BEC.

Recently, it was shown \cite{RR:PSV:PRE1} that one can manipulate
trajectories of bright solitons and their intensity distribution in the
coupled nonlinear Schr\"{o}dinger (NLS) equations by adjusting the
respective self-phase-modulation (SPM) and cross-phase modulation (XPM)
interactions. Experiments, starting from early works \cite{old}, and
including more recent ones \cite{new}, made use of the repulsive
interactions in multi-component BECs composed of $^{87}$Rb atoms to
demonstrate the possibility of producing long-lived multiple-condensate
states. Those results, and the availability of multi-component condensates
with attractive interactions \cite{Randy,Inguscio,attractiverubidium85},
suggest one to look for a mechanism for enhancing the stability of
multi-component BECs by engineering properties of bright solitons, which are
coherent condensates by themselves.

In this paper, we consider the dynamics of a binary condensate in
an expulsive time-dependent parabolic potential, governed by a
system of coupled Gross-Pitaevskii (GP) equations. The motivation
to study the properties of the condensates in expulsive potentials
stems from the fact that BECs are more stable in the confining
trap while they get compressed and are more unstable in the
expulsive trap. Expulsive parabolic potentials  are also used in
experiments for probing various dynamical properties of the
condensates \cite{expulsive}. We then study the evolution of the
condensates assuming that the scattering length can be made
time-dependent through the FR, so as to cast the system into an
integrable form. We thus observe that the condensates in the
expulsive temporarily transient harmonic potential may sustain
stability, while their counterparts in the time-independent
potential would rapidly decay. The analytical results are
confirmed by numerical simulations employing the split-step
Crank-Nicolson method. Furthermore, we demonstrate that the
addition of noise does not impact the stability of the condensates
in the expulsive transient potential.

\section{The model and the Lax pair}

Considering a two-component BEC, with equal atomic masses $m$ and attractive
interactions (such as the condensate composed of two hyperfine states of $%
^{7}$Li \cite{Randy} or $^{85}$Rb \cite{attractiverubidium85,Billam} atoms),
trapped in a parabolic potential, the mean-field evolution of the setting is
governed by coupled quasi-one-dimensional GP equations \cite{pethick},
written in a scaled form:

\begin{eqnarray}
i\frac{\partial \psi _{1}}{\partial t} &=&\left( -\frac{1}{2}\frac{\partial
^{2}}{\partial x^{2}}+b_{11}|\psi _{1}|^{2}+b_{12}|\psi _{2}|^{2}+\frac{%
\lambda _{1}^{2}}{2}x^{2}\right) \psi _{1},  \notag \\
i\frac{\partial \psi _{2}}{\partial t} &=&\left( -\frac{1}{2}\frac{\partial
^{2}}{\partial x^{2}}+b_{21}|\psi _{1}|^{2}+b_{22}|\psi _{2}|^{2}+\frac{%
\lambda _{2}^{2}}{2}x^{2}\right) \psi _{2},  \label{modeleq}
\end{eqnarray}%
where $\psi _{j}$ ($j=1,2$) is the macroscopic wave function of the $j$-th
component subject to the normalization condition $\int_{-\infty }^{+\infty
}|\psi _{1}|^{2}dx=1$ and $\int_{-\infty }^{+\infty }|\psi
_{2}|^{2}dx=N_{2}/N_{1}$. The interaction between the atoms is described by
the self-interaction coefficients, $b_{jj}=4n_{jj}N_{i}$/$l_{\perp }$, and
the interaction between different components is controlled by $%
b_{jk}=4n_{jk}N_{i}$/$l_{\perp }$, where $n_{jj}$ is the scattering lengths
of the $j$-th components, and $n_{jk}$ is the same for collisions between $j$
and $k$ . The dynamics of the two-component BECs composed of hyperfine
states of $^{87}$Rb atoms has been experimentally investigated in detail,
and it it was shown how the binary condensates generate various patterns \cite%
{old,new}. The system of coupled GP equations for the two-component BEC with
unequal scattering lengths is nonintegrable, and has been investigated
earlier \cite{unequalsl}. In the present work, we consider the condensates
with symmetric interaction strengths, \cite{Rajendren,twosol.vr.2010}
assuming $b_{11}=b_{21}=b_{12}=b_{22}$ (the scattering lengths in attractive
binary mixtures may be indeed made nearly equal \cite{Randy,scatt-length,Billam}),
and $\lambda _{j}^{2}(t)=\omega _{j}^{2}(t)/\omega _{\perp }^{2}$, where $%
\omega _{j}$ and $\omega _{\perp }$ represent the angular frequencies of the
trapping potential in the axial and radial directions, respectively. Time $t$
and coordinate $x$ are measured in units of $2/\omega _{\perp }$ and $%
l_{\perp }=\sqrt{\hbar /m\omega _{\perp }}$, respectively, while $l_{\perp }$
represents the linear oscillator length of the tight trapping potential
acting in the transverse direction.

Further, assuming $\omega _{1}(t)=\omega _{2}(t)=\omega (t)$ (i.e., $\lambda
_{1}=\lambda _{2}\equiv \lambda $), and allowing the scattering lengths $%
b_{jk}$ and the strength of trapping potential $\lambda ^{2}$ to vary with
time, eq. (\ref{modeleq}) takes the following form:

\begin{eqnarray}
i\psi _{1t}+\psi _{1xx}+2g(t)(|\psi _{1}|^{2}+|\psi _{2}|^{2})\psi
_{1}+\lambda ^{2}(t)x^{2}\psi _{1}=0  \notag \\
i\psi _{2t}+\psi _{2xx}+2g(t)(|\psi _{1}|^{2}+|\psi _{2}|^{2})\psi _{2}
+\lambda ^{2}(t)x^{2}\psi _{2}=0.  \label{gptwo}
\end{eqnarray}

\begin{figure}[tbp]
\begin{center}
\includegraphics[scale=0.4]{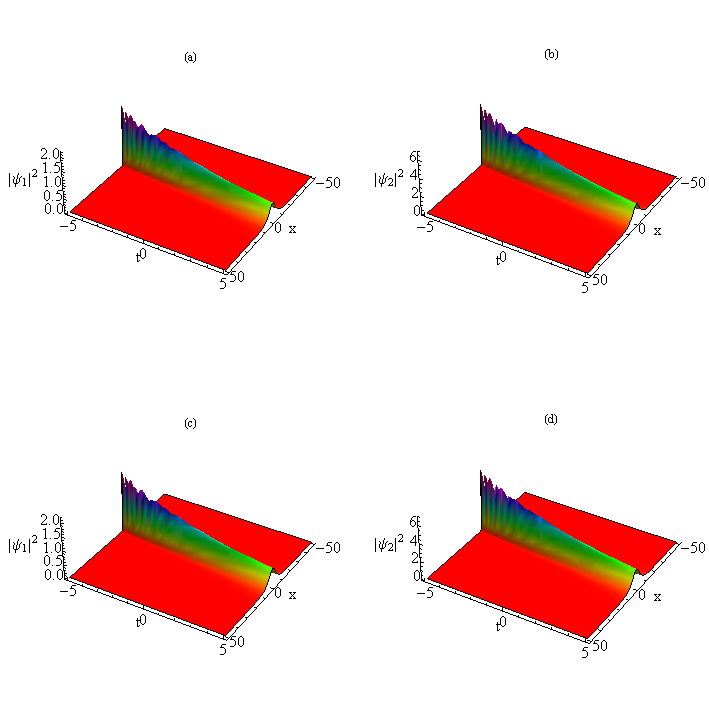}
\end{center}
\caption{\textbf{Upper Panels (a-b)}: The analytical density profile of the
bright solitons in the time-independent expulsive trap for $\Gamma =-0.25$
and $g(t)=0.5e^{-0.25t}$. \textbf{Lower Panels (c-d)}: The numerically
simulated density profile for the same case.}
\label{fig.ana1}
\end{figure}

It is relevant to stress, in passing, that we consider the GP equations in the framework
of the mean-field approximation, thus completely neglecting quantum correlations
which, beyond the limits of the mean-field theory, may cause macroscopic entanglement
between matter-wave solitons via their collisions \cite{Maciek,Billam,entanglement}.

In eqs. (\ref{gptwo}), the nonlinearity takes the Manakov's form \cite%
{Manakov} (which is a necessary condition if an integrable system is sought
for), with $g(t)$ standing for the common interaction strength, while $%
\lambda (t)$ represents the time-dependent trap frequency. Below,
we consider values $g(t)>0$, which correspond to the attractive
signs of intraspecies and interspecies interactions. Under the
special integrability condition imposed on $g(t)$ and $\lambda
(t)$, (see eq. (\ref{integ}) below), eqs. (\ref{gptwo}) admit a
representation in the form of an eigenvalue problem,

\begin{eqnarray}
\Phi _{x} &+&U\Phi =0,  \notag \\
\Phi _{t} &+&V\Phi =0,  \label{Lax}
\end{eqnarray}%
\label{lax} where $\Phi =(\phi _{1},\phi _{2},\phi _{3})^{T}$ and
\begin{equation}
U=\left(
\begin{array}{ccc}
i\zeta (t) & Q_{1} & Q_{2} \\
-Q_{1}^{\ast } & -i\zeta (t) & 0 \\
-Q_{2}^{\ast } & 0 & -i\zeta (t) \\
&  &
\end{array}%
\right) ,  \label{U}
\end{equation}%
\begin{equation}
V=\left(
\begin{array}{ccc}
v_{11} & v_{12} & v_{13} \\
v_{21} & v_{22} & v_{23} \\
v_{31} & v_{32} & v_{33} \\
&  &
\end{array}%
\right) ,  \label{V}
\end{equation}%
with
\begin{eqnarray}
v_{11} &=&-i\zeta (t)^{2}+i\Gamma (t)x\zeta (t)+\frac{i}{2}Q_{1}Q_{1}^{\ast
}+\frac{i}{2}Q_{2}Q_{2}^{\ast }  \notag \\
v_{12} &=&\Gamma (t)xQ_{1}-\zeta (t)Q_{1}+\frac{i}{2}Q_{1x}  \notag \\
v_{13} &=&\Gamma (t)xQ_{2}-\zeta (t)Q_{2}+\frac{i}{2}Q_{2x}  \notag \\
v_{21} &=&-\Gamma (t)xQ_{1}^{\ast }+\zeta (t)Q_{1}^{\ast }+\frac{i}{2}%
Q_{1x}^{\ast }  \notag \\
v_{22} &=&i\zeta (t)^{2}-i\Gamma (t)x\zeta (t)-\frac{i}{2}Q_{1}Q_{1}^{\ast }
\notag \\
v_{23} &=&-\frac{i}{2}Q_{2}Q_{1}^{\ast }  \notag \\
v_{31} &=&-\Gamma (t)xQ_{2}^{\ast }+\zeta (t)Q_{2}^{\ast }+\frac{i}{2}%
Q_{2x}^{\ast }  \notag \\
v_{32} &=&-\frac{i}{2}Q_{1}Q_{2}^{\ast }  \notag \\
v_{33} &=&i\zeta (t)^{2}-i\Gamma (t)x\zeta (t)-\frac{i}{2}Q_{2}Q_{2}^{\ast }
\notag
\end{eqnarray}%
\begin{eqnarray}
Q_{1} &=&\frac{1}{\sqrt{g(t)}}\psi _{1}(x,t)e^{i\Gamma (t)x^{2}/2}  \notag \\
Q_{2} &=&\frac{1}{\sqrt{g(t)}}\psi _{2}(x,t)e^{i\Gamma (t)x^{2}/2}
\notag
\end{eqnarray}%
The compatibility condition, $(\Phi _{x})_{t}=(\Phi _{t})_{x}$, leads to the
zero curvature equation $U_{t}-V_{x}+[U,V]=0$ which yields the integrable
system of coupled GP equations (\ref{gptwo}), provided that the spectral
parameter $\zeta (t)$ obeys the following nonisospectral condition:
\begin{equation}
\zeta (t)=\mu \exp \left[ -\int \Gamma (t)dt)\right] ,  \label{eq:mu}
\end{equation}%
where $\mu $ is a hidden complex constant and $\Gamma (t)$ is an arbitrary
function of time, which is related to the trapping frequency by the
following constraint:
\begin{equation}
\lambda ^{2}(t)=\Gamma ^{2}(t)-d\Gamma (t)/dt.  \label{trap}
\end{equation}%
Then, it follows from eq. (\ref{trap}) that frequency $\lambda (t)$ is
related to scattering length $g(t)$ through the integrability condition:

\begin{equation}
-g(t)d^{2}g(t)/dt^{2}+2\left[ dg(t)/dt\right] ^{2}-\lambda ^{2}(t)g^{2}(t)=0.
\label{integ}
\end{equation}%
Thus, the coupled GP equations (\ref{gptwo}) is completely
integrable if the trapping frequency, $\lambda (t)$, and
scattering length, $g(t)$, are subject to constraint
(\ref{integ}). For the time-independent trap, $\lambda
(t)=\mathrm{const}\equiv c_{1}$, eq. (\ref{integ}) yields
$g(t)=e^{c_{1}t}$ \cite{integref}. In this work, we focus on the
consideration of the integrable system satisfying condition
governed by eq. (\ref{integ}). If it is slightly broken, the
result depends on the accumulation of the deviation from the
integrability over the time interval, $T$, corresponding to
essential dynamical regimes, which may be easily identified in all
examples displayed below. Namely, if the deviation from the
integrability is characterized by difference $\Delta \lambda (t)$
from the value imposed by eq. (\ref{integ}), the condition for the
system to
remain close to the integrability is%
\begin{equation}
\int_{0}^{T}\Delta \lambda (t)dt\ll 1.  \label{<<}
\end{equation}

It should be mentioned that one can convert the coupled GP equation into the
celebrated Manakov's model by a suitable transformation, as shown, e.g., in
ref. \cite{Rajendren}. Nevertheless, the direct formulation of the
integrability formalism in the form of eqs. (\ref{U}), (\ref{V}), and (\ref%
{integ}) is quite useful, as it makes it possible to apply the gauge-transformation
method for generating multisoliton solutions, and it may also be used for the
search of more general integrable systems.

\begin{figure}[tbp]
\begin{center}
\includegraphics[scale=0.4]{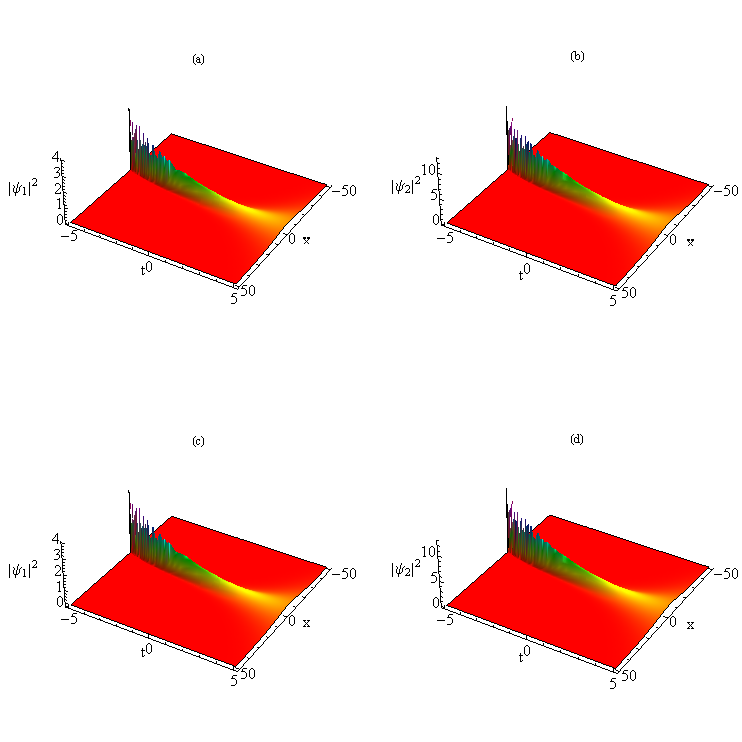}
\end{center}
\caption{\textbf{Upper panel (a-b): }Analytical results\textbf{\ }for the
compression of the condensate in the time-independent expulsive trap for $%
\Gamma =-0.5$ and $g(t)=0.5e^{-0.5t}$. \textbf{Lower Panel (c-d)}:
Numerically simulated density profile for the same case.}
\label{fig.ana2}
\end{figure}

\section{Analytical and numerical results for two-component bright solitons
in the integrable system}

Using the gauge-transformation approach \cite{llchaw1991}, bright solitons
of the coupled GP equation (\ref{gptwo}), subject to the integrability
condition (\ref{integ}), can be found in the form of

\begin{eqnarray}
\psi _{1}^{(1)} &=&\sqrt{\frac{1}{g(t)}}\varepsilon _{1}^{(1)}2\beta _{1}(t)%
\mathrm{sech}(\theta _{1})e^{i(-\xi _{1}+\Gamma (t)x^{2}/2)},
\label{coupledgponesol1} \\
\psi _{2}^{(1)} &=&\sqrt{\frac{1}{g(t)}}\varepsilon _{2}^{(1)}2\beta _{1}(t)%
\mathrm{sech}(\theta _{1})e^{i(-\xi _{1}+\Gamma (t)x^{2}/2)},
\label{coupledgponesol2}
\end{eqnarray}%
where
\begin{eqnarray}
\theta _{1} &=& 2 \beta _{1} x + 4\int \alpha _{1}\beta
_{1}dt-2\delta _{1},
\notag \\
\xi _{1} &=& 2 \alpha _{1} x + 2\int (\alpha _{1}^{2}-\beta
_{1}^{2})dt - 2\chi _{1}, \notag
\end{eqnarray}%
with $\alpha _{1}=\alpha _{10}\exp \left[ \int {\Gamma (t)}dt\right] $, $%
\beta _{1}=\beta _{10}\exp \left[ \int {\Gamma (t)}dt\right] $ while $\delta
_{1}$ and $\chi _{1}$ are arbitrary parameters, and $\varepsilon
_{1}^{(1)},\varepsilon _{2}^{(1)}$ are coupling coefficients, which are
subject to constraint $|\varepsilon _{1}^{(1)}|^{2}+|\varepsilon
_{2}^{(1)}|^{2}=1$. The gauge-transformation approach can be extended to
generate multi-soliton solutions \cite{twosol.vr.2010}.

\begin{figure}[tbp]
\begin{center}
\includegraphics[scale=0.4]{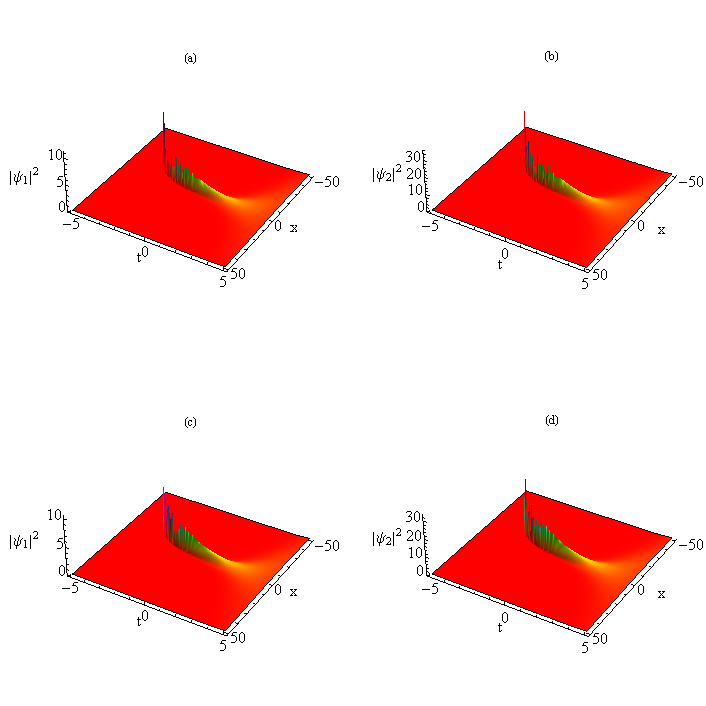}
\end{center}
\caption{\textbf{Upper panels (a-b)}: Onset of the decay (analytical
results) for $\Gamma =-0.9$ and $g(t)=0.5e^{-0.9t}$ in the expulsive
time-independent trap. \textbf{Lower Panels (c-d)}: Numerically simulated
density profile for the same case, showing the decay.}
\label{fig.ana3}
\end{figure}
Thus, it is obvious from the above that the amplitude of the bright solitons
is determined by the temporally modulated scattering length $g(t)$ and trap
frequency $\Gamma (t)$ ($\beta _{1}(t)$ varies exponentially with $\Gamma (t)
$).

To start the analysis of particular solutions relevant for the physical
realization of the system, we now switch off the time dependence of the
harmonic trap and adopt the scattering length in the form of $%
g(t)=0.5e^{-0.25t}$, for which eqs. (\ref{trap}) and (\ref{integ}) render
the parabolic potential expulsive, $\Gamma =-0.25$. The corresponding
density profile of the condensate is shown in the upper panel of fig. \ref%
{fig.ana1} (figs. \ref{fig.ana1}(a) and \ref{fig.ana1}(b)). The
corresponding density profile of the condensates produced by numerical
simulations of the real time propagation, using the split-step
Crank-Nicolson method is shown in figs. \ref{fig.ana1}(c) and \ref{fig.ana1}
(d). From figs. \ref{fig.ana1}(a)-\ref{fig.ana1}(d), we observe perfect
agreement between the analytical and numerical results. When we double the
trapping frequency to $\Gamma =-0.5$, keeping the trap expulsive and choose
the scattering length as $g(t)=0.5e^{-0.5t}$ consistent with eqs.(\ref{trap}
) and (\ref{integ}), the compression of the condensates sets in, as shown in
figs. \ref{fig.ana2}(a-b). This is confirmed by numerical simulations shown
in figs. \ref{fig.ana2}(c) and \ref{fig.ana2}(d). When we further enhance
the trap strength, to $3.6$ times the original value, $\Gamma =-0.9$,
keeping the trap expulsive and choose the scattering length as $%
g(t)=0.5e^{-0.9t}$ (again consistent with the integrability condition given
by eqs. (\ref{trap}) and (\ref{integ})), one observes decay (spreading out)
of the condensates, as shown in figs. \ref{fig.ana3}(a-b). Again, results of
numerical simulations, shown in figs. \ref{fig.ana3}(c) and \ref{fig.ana3}
(d), match with figs. \ref{fig.ana3}(a) and \ref{fig.ana3}(b), respectively.

\begin{figure}[tbp]
\begin{center}
\includegraphics[scale=0.4]{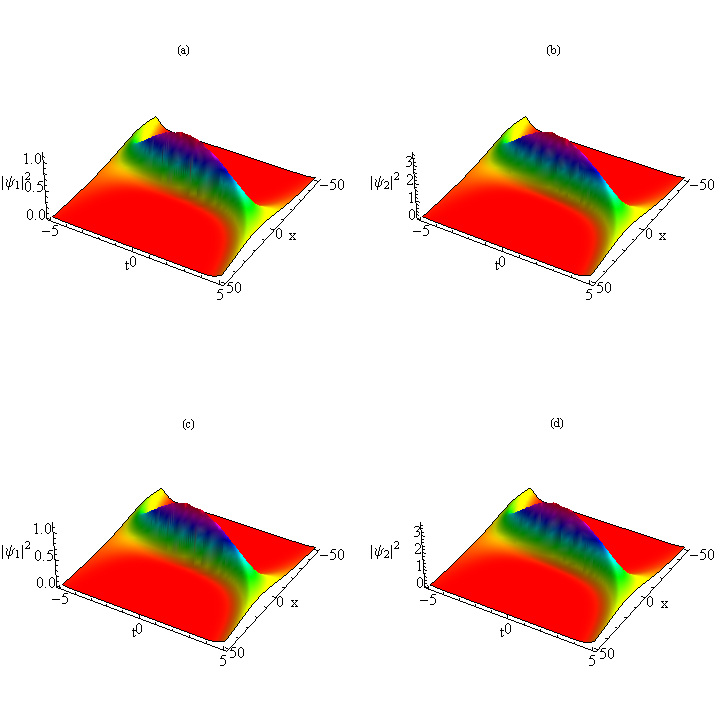}
\end{center}
\caption{\textbf{Upper panel (a-b)}: The analytical density profile produced
by switching on the time dependence of the expulsive trap for $\Gamma
(t)=-0.25t$ and $g(t)=0.5e^{-0.125t^{2}}$. \textbf{Lower Panel (c-d)}:
Numerically simulated density profile for $\Gamma (t)=-0.25t$ and $%
g(t)=0.5e^{-0.125t^{2}}$.}
\label{fig.ana4}
\end{figure}

\begin{figure}[tbp]
\begin{center}
\includegraphics[scale=0.4]{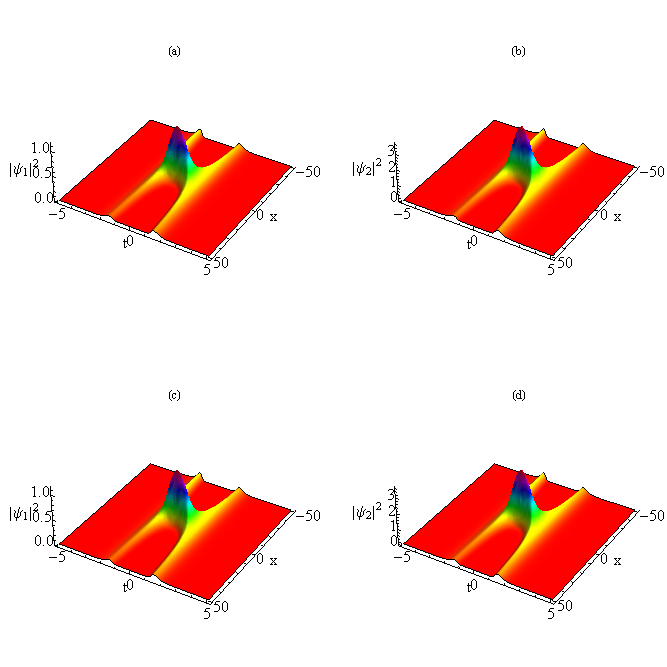}
\end{center}
\caption{\textbf{Upper panels (a-b)}: Analytically found density profiles of
the condensates in the time-dependent expulsive potential for $\Gamma (t)=-5t
$ and $g(t)=0.5e^{-2.5t^{2}}$. \textbf{Lower panels (c-d)}: Numerically
found density profiles for the same case.}
\label{fig.ana5}
\end{figure}

To enhance the stability of the condensates, we now switch ON the time
dependence of the harmonic trap, keeping it expulsive, as $\Gamma (t)=-0.25t$%
, and choose $g(t)=0.5e^{-0.125t^{2}}$, which is consistent with eqs. (\ref%
{trap}) and (\ref{integ}). The corresponding density profile is
shown in figs. \ref{fig.ana4}(a,b). This analytical solution is
confirmed by the numerical simulations, as shown in figs.
\ref{fig.ana4}(c,d). When we increase the expulsive-potential
strength by a factor of $20$ (as compared to
fig.(\ref{fig.ana4})), accordingly setting
$g(t)=0.5e^{-2.5t^{2}}$, which is consistent with eqs.
(\ref{trap}) and (\ref{integ}), the corresponding density profile
shown in figs. \ref{fig.ana5} (a,b) again matches with its
numerical counterpart shown in figs. \ref{fig.ana5}(c) and
\ref{fig.ana5}(d).

\begin{figure}[tbp]
\begin{center}
\includegraphics[scale=0.4]{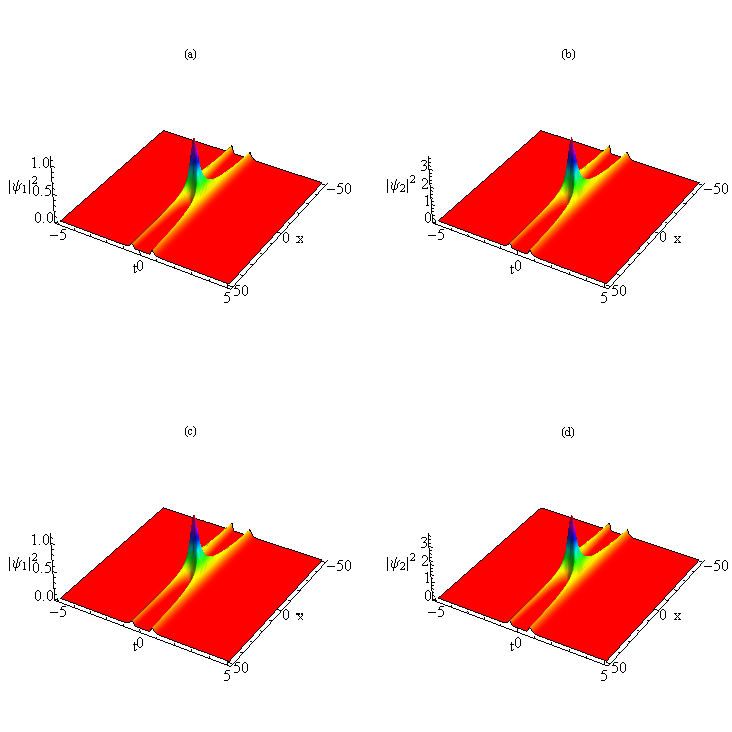}
\end{center}
\caption{\textbf{Upper panels (a-b)}: The analytically found density profile
in the time-dependent expulsive potential for $\Gamma (t)=-25t$ and $%
g(t)=0.5e^{-12.5t^{2}}$. \textbf{Lower panels (c-d)}: The numerically
simulated density profile for the same case. }
\label{fig.ana6}
\end{figure}

When we further increase the-time dependent expulsive-trap frequency $\Gamma
(t)$ by a factor of $100$ (as compared to fig.(\ref{fig.ana4})), and choose
the interaction strength in accordance with eqs.(\ref{trap}) and (\ref{integ}%
)), an abrupt increase in the density is no more observed, as figs. \ref%
{fig.ana6}(a-b) show. In fact, any further increase of $\Gamma (t),$ and the
respective change of the interaction strength $g(t)$ consistent with eqs. (%
\ref{trap}) and (\ref{integ}), does not lead to a significant increase in
the density, even though the attractive-interaction strength increases
rapidly. In other words, the condensates with the attractive interactions in
the time-dependent expulsive parabolic potential remain essentially stable
over a reasonably large interval of time. Again,analytical results match the
numerical simulations, as shown in figs. \ref{fig.ana6} (c-d).

Thus, we observe that the two-component BEC with attractive interactions in
the time-dependent expulsive trap, described by the integrable system, is more
long-lived, as its lifetime can
be increased by means of the FR management, in comparison with the
condensates in the time-independent expulsive trap. Note that the scalar
(single-component) condensate, stabilized by means of the Feshbach
management in a similar setting, quickly decays in the course of the
evolution.

\begin{figure}[tbp]
\begin{center}
\includegraphics[scale=0.45]{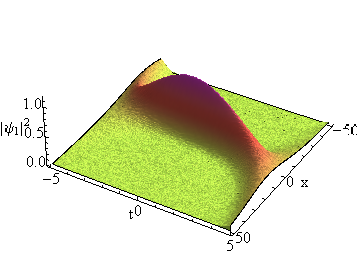} %
\includegraphics[scale=0.45]{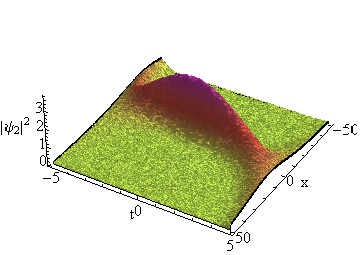}
\end{center}
\caption{The numerically simulated density profile, corresponding to fig.
\protect\ref{fig.ana4}, but withthe addition of random noise.}
\label{fig.nuwn4}
\end{figure}

To further confirm the stabilization, we added random noise to simulations
performed in the time-dependent expulsive trap. The respective numerically
simulated density profiles demonstrate, in figs. (\ref{fig.nuwn4})-(\ref%
{fig.nuwn6}), that the noise does not impact the stability of the
condensates. In all the above numerical simulations, the norm of
the wavefunctions is conserved. This conclusion corroborates that
the two-component condensate in the time-dependent expulsive
harmonic trap, governed by the specially devised integrable
system, is more stable in comparison with its counterpart in the
time-independent trap. The above results indicate the possibility
of increasing the life span of the two-component BEC with
attractive interactions in the time-dependent parabolic potential
experimentally, employing the FR management, which may be applied
to the condensates composed of $^{39}$K \cite{Inguscio}, $^{85}$Rb
\cite{attractiverubidium85} and $^{7}$Li \cite{Lithium-rubidium}
atoms.

\begin{figure}[tbp]
\begin{center}
\includegraphics[scale=0.45]{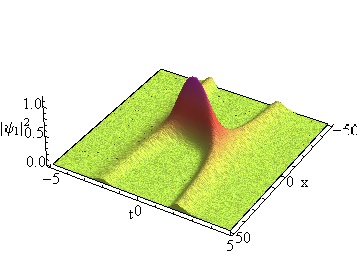} %
\includegraphics[scale=0.45]{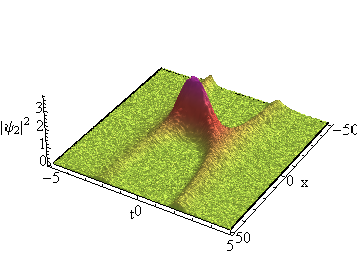}
\end{center}
\caption{The numerically simulated density profile, corresponding
to fig. (\ref{fig.ana5}), but with the addition of random noise.}
\label{fig.nuwn5}
\end{figure}

\begin{figure}[tbp]
\begin{center}
\includegraphics[scale=0.45]{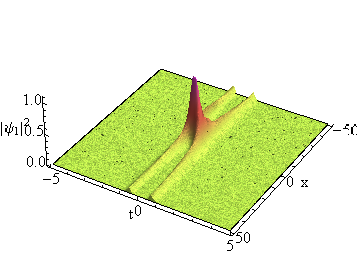} %
\includegraphics[scale=0.45]{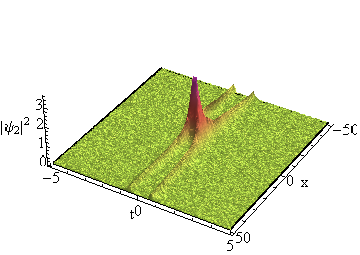}
\end{center}

\caption{The numerically simulated density profile, corresponding
to fig.\protect\ref{fig.ana6} but with     the addition of random
noise.} \label{fig.nuwn6}
\end{figure}

\section{Conclusion}

In this paper, we have shown that one can make use of the precise
control of the scattering length by means of the FR (Feshbach
resonance) to design conditions under which the mean-field
dynamics of two-component condensates in the time-dependent
expulsive parabolic potential is governed by the integrable model
(a deformation of the Manakov's system, with a time-dependent
scattering length). A regime may be selected, in which solitons
stay effectively stable against the decay for a reasonably large
interval of time, compared to the condensate in the
time-independent expulsive parabolic potential. The analytical
results, and their stability, have been corroborated by the
comparison with numerical simulations. In particular, it was
demonstrated that binary condensate with the attractive
interactions remains stable in the time-dependent harmonic
potential in spite of the addition of random noise. Thus, the
conclusion is that the bright solitons, which would quickly decay
(possibly following short-time compression) otherwise, may be made
more robust by the proper selection of scattering length through
FR management in  combination with the appropriate choice of the
time dependence of the parabolic potential.

\textbf{Acknowledgments}Authors thank the referees for their
suggestions to improve the contents of the paper. PSV and JBS
thank University Grants Commission (UGC) and Department of Science
and Technology (DST) (India), respectively, for the financial
support. RR acknowledges financial assistance received from DST
(Ref. No:SR/S2/HEP-26/2012), UGC (Ref. No: F.No 40-420/2011(SR),
Department of Atomic Energy -National Board for Higher Mathematics
(DAE-NBHM) (Ref.No: NBHM/R.P.16/2014/Fresh dated 22.10.2014) and
Council of Scientific and Industrial Research (CSIR) (Ref. No:
No.03(1323)/14/EMR-II dated 03.11.2014). This work was supported
by the NKBRSFC under grants Nos. 2011CB921502 and 2012CB821305,
and NSFC under grants Nos. 61227902 and 61378017.

\end{document}